\documentclass[traditabstract]{aa}
\usepackage{amsmath}
\usepackage{subfigure}
\usepackage{natbib}
\usepackage{amsmath}
\usepackage{multirow}
\usepackage{color}
\usepackage{longtable}
\usepackage{graphicx}
\usepackage{epstopdf}
\usepackage{booktabs}
\usepackage[colorlinks=true, linktocpage, linkcolor={blue!60!black}, citecolor={blue!60!black}, urlcolor={blue!60!black}]{hyperref}
\usepackage[dvipsnames,svgnames,x11names]{xcolor}

\usepackage{txfonts}

\begin{document}

\title{Validation of millimetre and submillimetre atmospheric collision induced absorption at Chajnantor
  \thanks{This publication is based on data acquired between October 2020 and September 2022 with the
    Atacama Pathfinder EXperiment (APEX) under programme ID 105.2084 within the framework of an ESO/ALMA development study under contract CFP/ESO/19/25417/HNE. APEX is a collaboration between the Max-Planck-Institut f\"ur
    Radioastronomie, the European Southern Observatory, and the Onsala Space Observatory.}}

\author{J.~R.~Pardo\inst{1,5}, C. De Breuck\inst{2}, D. Muders\inst{3}, J. Gonz\'alez\inst{2}, 
  J. P. P\'erez-Beaupuits\inst{4,3,7}, J. Cernicharo\inst{1}, C. Prigent\inst{5}, E. Serabyn\inst{6}, F.~M.~Montenegro-Montes\inst{2,8}
  T.~Mroczkowski\inst{2}, N. Phillips\inst{2} and E. Villard\inst{2} }

\institute{Consejo Superior de Investigaciones               
  Cientif\'{\i}cas, Instituto de F\'{\i}sica Fundamental,    
  Calle Serrano 121, 28006 Madrid, Spain \\                        
  \email{jr.pardo@csic.es}                                   
  \and    European Southern Observatory, Karl-Schwarzschild-Str. 2, 85748 Garching, Germany 
  \and    Max-Planck-Institut f\"ur Radioastronomie, Auf dem Hügel 69, 53121 Bonn, Germany 
  \and European Southern Observatory, Alonso de C\'ordova 3107, Vitacura Casilla 7630355, Santiago, Chile 
  \and Centre National de la Recherche Scientifique, Observatoire de Paris, Laboratoire d'Etudes du Rayonnement et la Mati\`ere en Astrophysique, 77 Avenue Denfert Rochereau, 75014 Paris, France.
  \and Jet Propulsion Laboratory, California Institute of Technology, 4800 Oak Grove Drive, Pasadena, CA 91109, USA
  \and Centro de Astro-Ingenier{\'i}a UC, Instituto de Astrof{\'i}sica, Pontificia Universidad Cat{\'o}lica de Chile, Avda Vicuña Mackenna 4860, Macul, Santiago Chile
  \and Departamento de F{\'i}sica de la Tierra y Astrof{\'i}sica e Instituto de F{\'i}sica de Part{\'i}culas y del Cosmos (IPARCOS). Universidad Complutense de Madrid, Ciudad Universitaria, 28040 Madrid, Spain
}

\date{Received MMMMMM DD, 2024; accepted MMMMM DD, 2024}

\abstract{
  Due to the importance of a reference atmospheric radiative transfer model for both planning and calibrating ground-based
  observations at millimetre and submillimetre wavelengths, we have undertaken a validation campaign consisting of
  acquiring atmospheric spectra under different weather conditions, in different diurnal moments and seasons, with the Atacama
  Pathfinder EXperiment (APEX), due to the excellent stability of its receivers and the very high frequency
  resolution of its backends. As a result, a data set consisting of 56 spectra within the 157.3-742.1 GHz frequency range,
  at kHz resolution (smoothed to $\sim$2-10 MHz for analysis), and spanning one order of magnitude ($\sim$ 0.35-3.5 mm)
  in precipitable water vapour columns, has
  been gathered from October 2020 to September 2022.
  These data are unique for their quality and completeness and, due to the proximity of
  APEX to the Atacama Large Millimetre/Submillimetre Array (ALMA), they provide an excellent opportunity to validate the
  atmospheric radiative transfer model currently installed in the ALMA software. The main issues
  addressed in the 
  study are possible missing lines in the model, line shapes, vertical profiles of atmospheric physical 
  parameters and molecular abundances, seasonal and diurnal variations and collision induced absorption {(CIA)}, to which
  this paper is devoted, in its N$_2$-N$_2$ + N$_2$-O$_2$ + O$_2$-O$_2$ (dry), and N$_2$-H$_2$O + O$_2$-H$_2$O
  (``foreign'' wet) mechanisms. {All these CIA terms should remain unchanged in the above mentioned ALMA atmospheric
    model as a result of this work.} }


\titlerunning{Validation of mm/submm atmospheric model at Chajnantor with APEX}
\authorrunning{J.~R.~Pardo et al.}    

\maketitle

\section{Introduction}
\label{sct:intro}

Observations of cold dark clouds, star-forming regions, evolved stars, galaxies, and other objects in space, 
through millimetre and submillimetre wavelength observations have broadened our vision of the Universe over
the last few decades. These observations are usually performed from high and dry sites,
where conditions
allowing some sky transparency for them are found, as \cite{Hills1978} revealed with pioneering site testing
observations from Tenerife at $\sim$2400 meters above sea level. Even so, the atmospheric millimetre
and submillimetre spectrum is rather complicated and fast changing, both in space and time. Site testing for
ground-based submillimetre observatories and field spectroscopy and radiometry experiments continued over the
years in the Atacama desert (Chile), Hanle (India), the Tibet Plateau,
Greenland, the South Pole, Mauna Kea (Hawai'i), and a few other sites. The work presented here is a continuation
of the Mauna Kea studies, \cite{Pardo2001a} and \cite{Pardo2005}, as it uses their resulting atmospheric radiative
transfer model \citep{Pardo2001b}
as the starting point. For a wider view on the subject, see a 2012 review in \cite{Tremb2012} and
several complementary publications from the last 25 years, such as \cite{Mats1999}, \cite{Paine2000}, 
\cite{Cham2003}, \cite{Ji2004}, {\cite{Shi2016}}, \cite{Mats2017}, \cite{Mlaw2019}, \cite{Ning2020}, \cite{Pardo2022} and
\cite{Mlaw2023}.

\begin{figure*}[h] 
 	\centering 
	\includegraphics[width=1.00\textwidth]{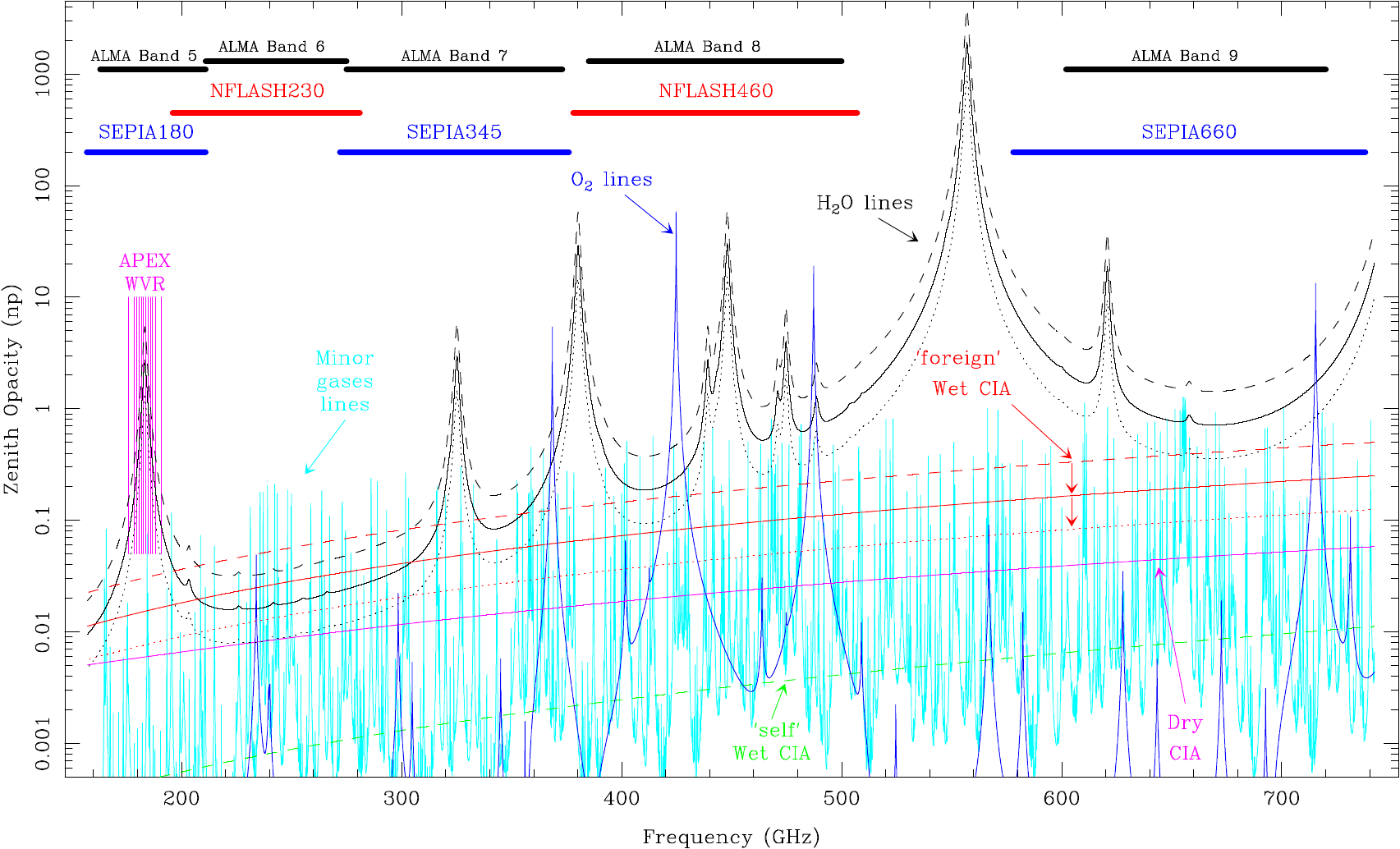}                              
	\caption{Reference ATM model for the APEX site, considering 555 hPa and 273 K as physical
          parameters at the ground, and 0.5 (dotted black and red lines), 1.0 (solid black and
          red lines) and 2.0 mm (dashed black and red lines) of precipitable water vapour column {(PWVC)}.
          Note that the H$_2$O-H$_2$O {(self wet)} CIA is in practice negligible at these very dry conditions
          ({only the curve for 2.0 mm PWVC is shown in dashed green color as} the curves
          for 1.0 and 0.5 mm PWVC are below the bottom {Y limit} of the plot). 
          The frequency ranges covered by the SEPIA and nFLASH receivers used in this work are
          plotted for reference, as well as the ALMA bands 6 to 9. The central frequencies of the
          6 double side-band channels of the APEX Water Vapour Radiometer, on both sides of the
          183.31 GHz H$_2$O line, are also plotted. The red and pink CIA curves are the ones under
          validation in this paper. {Cyan (minor gases opacity), blue (O$_2$ lines opacity)
          and pink (dry CIA opacity) curves remain unchanged for all PWVC values.}}
        \label{fg:atmref}
\end{figure*}

The contributors to the millimetre and submillimetre atmospheric spectrum are O$_2$ 
through magnetic dipolar (M1) rotational transitions, H$_2$O through electric dipolar (E1)
rotational transitions, weaker E1 features from other ``minor'' atmospheric gases,
such as N$_2$O, CO, HCN, HCl, O$_3$ and its isotopologues and vibrationally excited states, and, finally,
non-resonant collision-induced absorption
(CIA) due to several mechanisms: N$_2$-N$_2$, O$_2$-O$_2$ and O$_2$-N$_2$
collisions (dry CIA), and H$_2$O-N$_2$ + H$_2$O-O$_2$ collisions (“foreign” wet CIA),
as well as H$_2$O-H$_2$O collisions (“self” wet CIA). This last term is almost two orders of magnitude
lower than the “foreign” CIA in the dry conditions ({precipitable water vapor column,}
PWVC < $2\,\rm{mm}$) relevant to observations
at frequencies above approximately 170 GHz, and, therefore, does not play a role in this study.

The importance of a reference atmospheric radiative transfer model for both planning and
contributing in the calibration of ground-based
observations at millimetre and submillimetre wavelengths is the main motivation of this study. 
From 2006 to 2011 the C++ implementation of \cite{Pardo2001b} atmospheric
transmission model ATM was achieved, under European Southern Observatory (ESO) contract
14977/ESO/07/15694/YWE, to the Telescope Calibration 
subsystem (TelCal) of the official ALMA software. ATM includes the spectroscopy and reference vertical
profiles for all relevant molecular species contributing to the millimetre and submillimetre
atmospheric spectrum as seen from ground-based observatories, along with empirical and
theoretical descriptions of the CIA mechanisms. The software was delivered on schedule
and has been used within ALMA since the first cycle of science observations. Figure \ref{fg:atmref} shows a
reference calculation of the different
opacity terms in ATM under typical conditions at the APEX site in the Chajnantor Plateau, assuming
internal model parameters
as in Section \ref{sct:analysis}. The dry opacity component (dark blue, light blue and pink lines in the
figure) is quite constant at the site, with maximum variations of just a few percent. However, the wet
part of the opacity (black and red lines in the figure) has a dynamic range of more than one order of
magnitude, even in clear sky conditions, and significant changes can be expected at time scales as short
as a few minutes. 

\begin{figure*}[h ]
 	\centering 
	\includegraphics[width=0.95\textwidth]{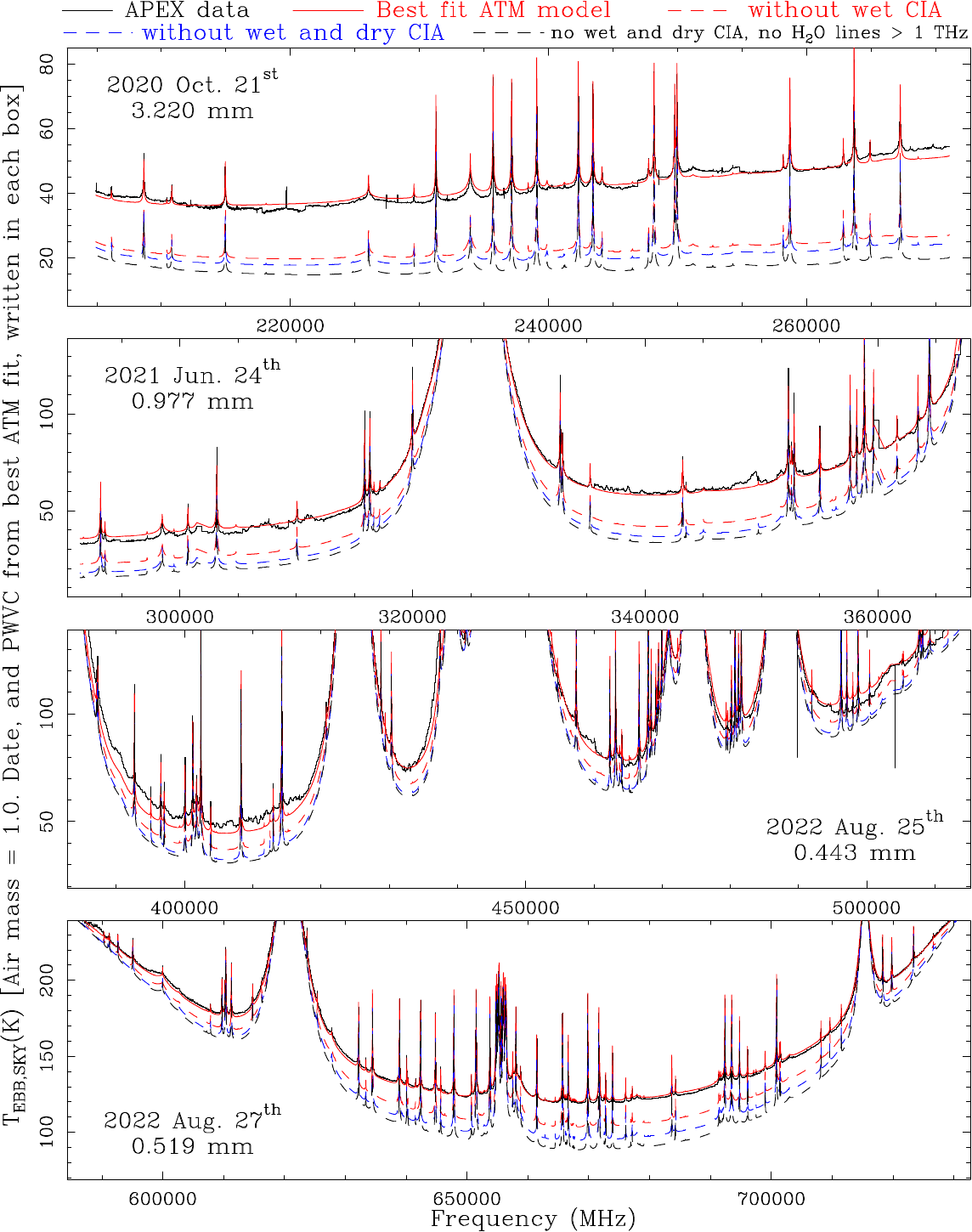}           
	\caption{Selection of spectra used in this study, showing the result of the best PWVC fit given by the ATM full model,
          and the effect on the atmospheric brightness temperature when the same fitted PWVC is used but different opacity contributions
          are removed from the calculations.}
        \label{fg:APEXRetrievalTestComp}
\end{figure*}

Figure \ref{fg:atmref} shows the importance of CIA absorption for the total atmospheric opacity and brightness
temperature. For example, the ``foreign'' wet CIA exceeds the H$_2$O lines opacity for most of the $\sim$205-302 GHz 
range and this translates into the large effect of removing it in the model seen in Figure
\ref{fg:APEXRetrievalTestComp}. CIA terms also explain that the atmospheric transmission in submillimetre
windows is not as good as a simple ``lines only'' model would predict. For this reason, the validation
of atmospheric CIA terms is so important for millimetre and submillimetre observatories, and is 
the main motivation of this paper. 

Besides astronomy, the study of CIA terms is also of interest for Earth observations and the
remote sensing community. Current operational meteorological applications are limited to
200 GHz. The upcoming EUropean organisation for the exploitation of METeorological SATellites
(EUMETSAT) Polar System-Second Generation (MetOp-SG), to be launched in 2025, will carry
an instrument, the Ice Cloud Imager (ICI), with frequencies up to 664 GHz \citep{kl2016}. The
main objective of ICI is to provide data on humidity and ice hydrometeors, particularly the
bulk ice mass that can be uniquely quantified with millimeter wave observations
from satellites. Furthermore, the deployment of the EUMETSAT Polar System – Sterna
(EPS-Sterna) constellation, consisting of 6 micro-satellites for launch around 2030, will
enhance Numerical Weather Prediction (NWP) accuracy.  Each micro-satellite will carry a
sounder, including channels around the H$_2$O line at 325 GHz, providing global
temperature and water vapor profiles with unparalleled coverage and revisit time.
The fast radiative transfer models
used for satellite data assimilation in Numerical Weather Prediction (NWP) models
are based on atmospheric line-by-line spectroscopy and on the so-called
MT-CKD (Mlawer-Tobin-Clough-Kneizys-Davies)
water vapour CIA (\cite{Mlaw2019}, \cite{Mlaw2023}). The MT-CKD is primarily tuned for
Earth energy budget, and as a consequence focuses mainly on frequencies higher
than 3 THz, as frequencies below are only weakly contributing to the surface or
top-of-atmosphere energy budgets. The carefully calibrated
APEX observations presented in this work provide additional constraints to the water vapour CIA, to
validate existing models in the millimetre and submillimetre wave ranges.

Most site testing and model validation campaigns at millimetre and submillimetre wavelengths have
been carried out with broadband (several hundreds of GHz) Fourier Transform Spectrometers (FTSs),
with frequency resolutions ranging from $\sim$0.2 to 10 GHz, and tipping radiometers using ``window''
frequencies such as 220, 225, 492 or 850 GHz (see previous references in this section). Our study
is the first to offer broadband coverage (several hundred of GHz) and very high spectral resolution
(better than 1 MHz), although this last aspect is not critical for CIA validation and we have smoothed
the resolution to $\sim$10 MHz. The original resolution,
however, will be preserved in future papers in order to focus on the fine details of the atmospheric spectrum such
as line shapes and very weak spectral features.  Section \ref{sct:ins} presents the telescope, receivers
and the instrumental set-up. A crucial update on the calibration, with respect to
our previous \cite{Pardo2022} work, can be found in Section \ref{sct:cal}. The observing runs
are described in Section
\ref{sct:obs} along with the useful spectra that will be used for the detailed analysis presented
in Section \ref{sct:analysis}. The output from this analysis leads to a discussion in Section \ref{sct:discussion}
that will focus on the role of CIA, and far-wings of
supra-THz H$_2$O lines, in the atmospheric spectrum. Summary and conclusions are given in Section
\ref{sct:concl}.

\section{Instrumental set-up}
\label{sct:ins}

The Atacama Pathfinder EXperiment {(APEX)} \citep{Gusten2006} is a world-class millimetre and
submillimetre observatory operating at a distance of roughly 2 km from the centre
of ALMA, at the Chajnantor plateau, on the Chilean Andes, 5105 meters above sea level.
It hosts a large variety of instruments, from several European partners, due to the
experimental nature of this project. The APEX antenna has a diameter of 12 meters and is
made of 264 aluminium panels. The full surface
accuracy is better than 15 $\mu$m r.m.s. Among the different instruments available, there are
five sideband separating (2SB) heterodyne (Het) superconductor–insulator–superconductor (SIS) tunnel
junction receivers that have been used for this project, connected to Fast Fourier Transform
Spectrometers (FFTS) providing kHz spectral resolutions. The good performance of them for our goals, and
the proximity to ALMA, provide an excellent opportunity
for validating the atmospheric model used in the latter facility, but also in APEX and
other millimetre and submillimetre observatories around the world. 

A first set of receivers, called SEPIA for Swedish ESO PI Instrument at APEX, are in a cryostat that
can accommodate 3 ALMA-like receiver cartridges with 
tertiary optics to illuminate them inside the Nasmyth cabin A of the APEX telescope. They 
were designed, constructed, and installed by the Group for Advanced Receiver Development (GARD) at
Onsala Space Observatory (OSO) in Sweden. The SEPIA cryostat was also manufactured by the GARD team. A
complete technical description can be found in \cite{Bel2018}, for SEPIA180 and SEPIA660, and
\cite{mele2022} for SEPIA345. SEPIA660 was used in our previous \cite{Pardo2022} publication. 

\begin{itemize}
\item{SEPIA180 (159-211 GHz)} is a dual polarisation 2SB receiver built to the specifications
  of ALMA Band 5 (it is based on the pre-production version of such receiver). It has two intermediate
  frequency (IF) outputs per polarisation, upper/lower sideband (USB/LSB), each covering 4-8 GHz, adding
  up a total of 16 GHz instantaneous IF bandwidth.
  The central frequencies of the two sidebands are separated by 12 GHz. The sideband rejection ratio is by
  design >10 dB and 18.5 dB on average. The single-sideband noise temperature is below 55 K at all frequencies
  within the band.\\
  
  \item{SEPIA345 (272–376 GHz)} is a dual polarisation 2SB receiver delivered in 2020. It has
    two IF outputs per polarisation, USB and LSB, each covering 4-12 GHz, adding up a total of 32 GHz
    instantaneous IF bandwidth. The central frequencies of the two sidebands are separated by 16 GHz.
    Each sideband (and polarisation) is recorded by two FFTS spectrometer units, each of them sampling
    4 GHz in the following configuration: FFTS1: 4.17 – 8.17 GHz IF bandwidth, FFTS2: 8.07 – 12.07 GHz
    IF bandwidth. Therefore, both units overlap in the middle for about 100 MHz and the full coverage is slightly
    smaller than 8 GHz (7.9 GHz). The sideband rejection ratio is by design >10 dB over 90\% of the band.
    Typical values of receiver temperature are $<$100 K below 320 GHz, approach 150 K at 340 GHz and then rise
    toward the high-frequency end of the receiver band.\\
    
  \item{SEPIA660 (597–725 GHz)} is a dual polarisation 2SB receiver that was installed and commissioned during
    the second half of 2018. The rest of the characteristics are as for SEPIA345 with the exception of the
    receiver temperature that is below 350 K at all frequencies within the band, and below 250 K in
    the central part of it. 

\end{itemize}

\begin{table*}[hbt!]
  \centering
  \setlength{\tabcolsep}{6.00pt} 
  \begin{tabular}{cccccccc}
    YYYY-    &   APEX       & Freq. & Air  & UT    & $\overline{P}_{gr}$(hPa) & $\overline{T}_{gr}$(K) & Figure \\
    MM-DD     & Receiver & (GHz) & Mass & (hh:mm)   & logfile & logfile & number  \\
  \hline 2020-10-21 & nFLASH230 & 204.9-271.1 & 1.00, 1.25, 1.50, 1.75, 2.00  & 06:42 to 08:00   & 554.7 & 273.0 & \ref{fg:APEXRetrievalTestComp}, \ref{fg:corr} \\  
  \hline 2020-12-05 & SEPIA345 & 274.8-343-1 & 1.00, 1.25, 1.50, 1.75, 2.00   & 01:05 to 02:59   & 555.1 & 269.5 & \ref{fg:corr}\\
  \hline 2020-12-06 & SEPIA660 & 582.1-742.1 & 1.00, 1.25, 1.50, 1.75, 2.00   & 10:45 to 13:55 & 555.0 & 273.1 & \ref{fg:20201206660}, \ref{fg:corr} \\     
  \hline 2021-06-24 & SEPIA345 & 291.0-371.0 & 1.41   & 20:50 to 21:08 & 553.7 & 260.8 &  \ref{fg:APEXRetrievalTestComp}, \ref{fg:corr} \\
  \hline 2021-10-31 & SEPIA180 & 157.3-205.2 & 1.00, 1.50, 2.00   &  23:41to 00:44  & 556.2 & 271.0 &  \ref{fg:cia_datasets}, \ref{fg:corr} \\
  2021-10-31 & nFLASH230 & 195.9-271.1 & 1.00, 1.50, 2.00   &  00:48 to 01:24  & 556.3 & 270.8 &  \ref{fg:cia_datasets}, \ref{fg:corr} \\
  2021-10-31 & SEPIA345 & 270.9-371.5 & 1.00, 1.50, 2.00   &  01:28 to 02:22  & 556.4 & 270.0 &  \ref{fg:corr} \\
  2021-10-31 & nFLASH460 & 453.0-518.0 & 1.00, 1.50, 2.00   &  02:38 to 04:58  & 556.3 & 272.1 &  \ref{fg:cia_datasets}, \ref{fg:corr} \\
  \hline 2022-05-11 & SEPIA660 & 582.0-739.0 & 1.00, 2.00   & 11:18 to 12:30   & 555.6 & 269.5 &  \ref{fg:corr} \\
  \hline 2022-08-25 & SEPIA660 & 582.0-739.0 & 1.00, 1.50, 2.00   & 16:37 to 17:59   & 555.3 & 272.8 &  \ref{fg:corr} \\
  2022-08-25 & nFLASH460 & 382.8-518.2 & 1.00   & 18:03 to 19:05   & 555.1 & 272.9 &  \ref{fg:corr} \\
  2022-08-25 & nFLASH460 & 382.8-517.1 & 1.00, 1.50   & 19:07 to 20:66   & 554.9 & 272.3 &  \ref{fg:cia_datasets}, \ref{fg:corr} \\
  2022-08-25 & SEPIA345 & 270.9-371.1 & 1.00, 1.50, 2.00   & 20:57 to 21:59   & 555.0 & 271.4 &  \ref{fg:corr} \\
  2022-08-25 & S180/N230/S345 & 174.8-371.1 & 1.00   & 21:41 to 22:42   & 555.1 & 270.3 & \ref{fg:APEXRetrievalTestComp}, \ref{fg:corr} \\
  \hline 2022-08-27 & SEPIA660 & 582.0-739.0 & 1.00, 1.50, 2.00   & 02:07 to 02:43   & 555.1 & 265.5 & \ref{fg:APEXRetrievalTestComp}, \ref{fg:corr} \\ 
  \hline 2022-08-28 & SEPIA180 & 158.8-211.2 & 1.00, 1.50, 2.00   & 14:19 to 14:37   & 556.3 & 273.0 &  \ref{fg:cia_datasets}, \ref{fg:corr} \\
  2022-08-28 & nFLASH230 & 195.9-271.1 & 1.00, 1.50, 2.00   & 15:14 to 15:26   & 556.1 & 273.6 &  \ref{fg:cia_datasets}, \ref{fg:corr} \\
  2022-08-28 & SEPIA345 & 270.9-371.1 & 1.00, 1.50, 2.00   & 15:53 to 17:00   & 556.0 & 273.6 &  \ref{fg:corr} \\
  2022-08-28 &  S180/N230 & 158.8-271.1 & 1.00   & 17:02 to 17:32   & 555.8 & 273.2 &  \ref{fg:corr} \\
  \hline 2022-09-01 & SEPIA660 & 587.9-649.1 & 1.00   & 16:38 to 16:46   & 551.9 & 268.8 &  \ref{fg:corr} \\
  2022-09-01 & SEPIA660 & 587.9-649.1 & 1.00   & 17:06 to 17:18   & 551.9 & 269.3 &  \ref{fg:corr} \\
  \hline 2022-09-03 & SEPIA660 & 587.9-732.0 & 1.00   & 16:02 to 16:24   & 553.0 & 269.5 &  \ref{fg:corr} \\
  \hline
  & & & & & & & \\
  \end{tabular}
  \caption{Summary of 56 atmospheric spectra, from 11 different observing runs, obtained with APEX from October 2020 to September 2022, that have been analysed in this work. Most of the
    observations were achieved by keeping constant the airmass and doing all the frequency tunings and scans before moving the telescope to
    a different elevation. A few observations, however, were done by keeping 
    the same frequency tuning and moving the telescope to take a scan for each elevation before tuning again. Since the atmospheric conditions evolve during the
    observations, neither observing mode can provide a view of a complete skydip under constant conditions, but the overall results provide a good consistency
    between data and model (see text for details).}
  \label{tb:retrievals}
\end{table*}

The other bands (nFLASH) were delivered in 2020 by the MPIfR Sub-mm technology division in Bonn. It is a receiver  
with two independently tunable frequency channels: nFLASH230 and nFLASH460. Both channels are dual polarisation
(2 SIS mixers per channel) and sideband separating (2SB) meaning 4 SIS junctions in total. The instrument is designed
to work as a dual colour receiver to allow simultaneously observing in both channels.

\begin{itemize}
\item{nFLASH230 (188-282 GHz)} has an extended IF coverage, from 4 to 12 GHz, and
  therefore it covers up to a total of 32 GHz IF instantaneous bandwidth including both sidebands and
  polarisations. The separation between the centre of the two sidebands is 16 GHz. Each sideband
  (and polarisation) is recorded by two spectrometer processors units (FFTS), each of them
  recording 4 GHz in the following configuration: FFTS1: 4.17 – 8.17 GHz IF bandwidth,
  FFTS2: 8.07 – 12.07 GHz IF bandwidth. Therefore, both units overlap in the middle for about 100 MHz
  and the full coverage is slightly smaller than 8 GHz (7.9 GHz). The typical receiver temperature
  is 60-80 K, increasing up to 80-90 K at the extremes of the frequency window (below 210 GHz or
  above 260 GHz LO frequencies). The sideband rejection is typically around 15 dB.\\

\item{nFLASH460 (377-508 GHz)} has 4-8 GHz output IF bandwidth,
  which is half the bandwidth that is covered by the nFLASH230 channel. The separation between the centre of
  the two sidebands is 12 GHz and each sideband (and polarisation) is covered by one FFTS spectrometer
  unit of 4 GHz bandwidth. The receiver temperatures are typically below 150 K, except at the high
  frequency end of the spectral window (LO frequency > 480 GHz) where this increases to higher values.
  Sideband rejection is typically better than 15 dB all over the band.

  \end{itemize}

The large sideband rejection ratio in all these bands is of the highest importance in this study.
Nevertheless, as atmospheric lines appear everywhere in the spectrum, we included frequency overlaps between adjacent tunings in our observational setup, to avoid systematics like residual contamination from the sidebands.
Further, we carefully checked for artifacts that may appear from the image band and found such issues to have a very limited impact on the final stitched spectra.

{The heterodyne receivers mentioned above are connected to
modern digital Fast Fourier Transform spectrometers (FFTS),
see \cite{kl2012}. They are based on high-speed analog-to-digital
converters (ADCs) and highly complex field-programmable gate array
(FPGA) chips for signal processing.
The FFTS used for these measurements transforms 4 GHz
instantaneous bandwidth into 64k spectral channels.
A 4-tap polyphase filter bank (PFB) algorithm is implemented
with a pipelined Fast Fourier Transform for continuous transformation
without any time gaps. Based on the PFB coefficients, this results
in an  equivalent noise bandwidth of 70.8 kHz, which is only
16\% wider than the frequency spacing.}
The spectral resolution is the main improvement, by more than
three orders of magnitude, with respect to our previous work
conducted $\sim$20-25 years ago from Mauna Kea in Hawai’i \citep{Pardo2001a},
{for which the finest spectral resolution was {$\sim$ 200 MHz}, clearly not
enough to resolve the spectral lines of many minor gases.}


\section{Calibration}
\label{sct:cal}
The final products of the observations presented in this work 
are atmospheric spectra in the form of Equivalent Blackbody Temperature as a function of frequency, T$_{\rm EBB}$($\nu$).
If F$_{atm}$($\nu$) is the calibrated atmospheric
flux at frequency $\nu$, T$_{\rm EBB}$($\nu$) is derived from: F$_{atm}$($\nu$)=B[T$_{\rm EBB}$($\nu$)]
with B being the blackbody function. In order to get F$_{atm}$($\nu$) or
T$_{\rm EBB}$($\nu$) free from the optical-electrical functions of the observing
system, two black bodies at different temperatures are observed with the receiver
(\citealt{UlichHaas76}; \citealt{Ulich80}). These black bodies are
implemented using a microwave absorber material, one at the receiver cabin
temperature (T$_{\rm{hot}}$), and the other one 
at a temperature near that of liquid N$_2$ at 5105~m altitude ($\approx$73~K).
The second absorber is installed in a small dewar and connected to a closed
cycle cryocooler.

\begin{figure*}[hbt!] 
 	\centering 
	\includegraphics[width=0.985\textwidth]{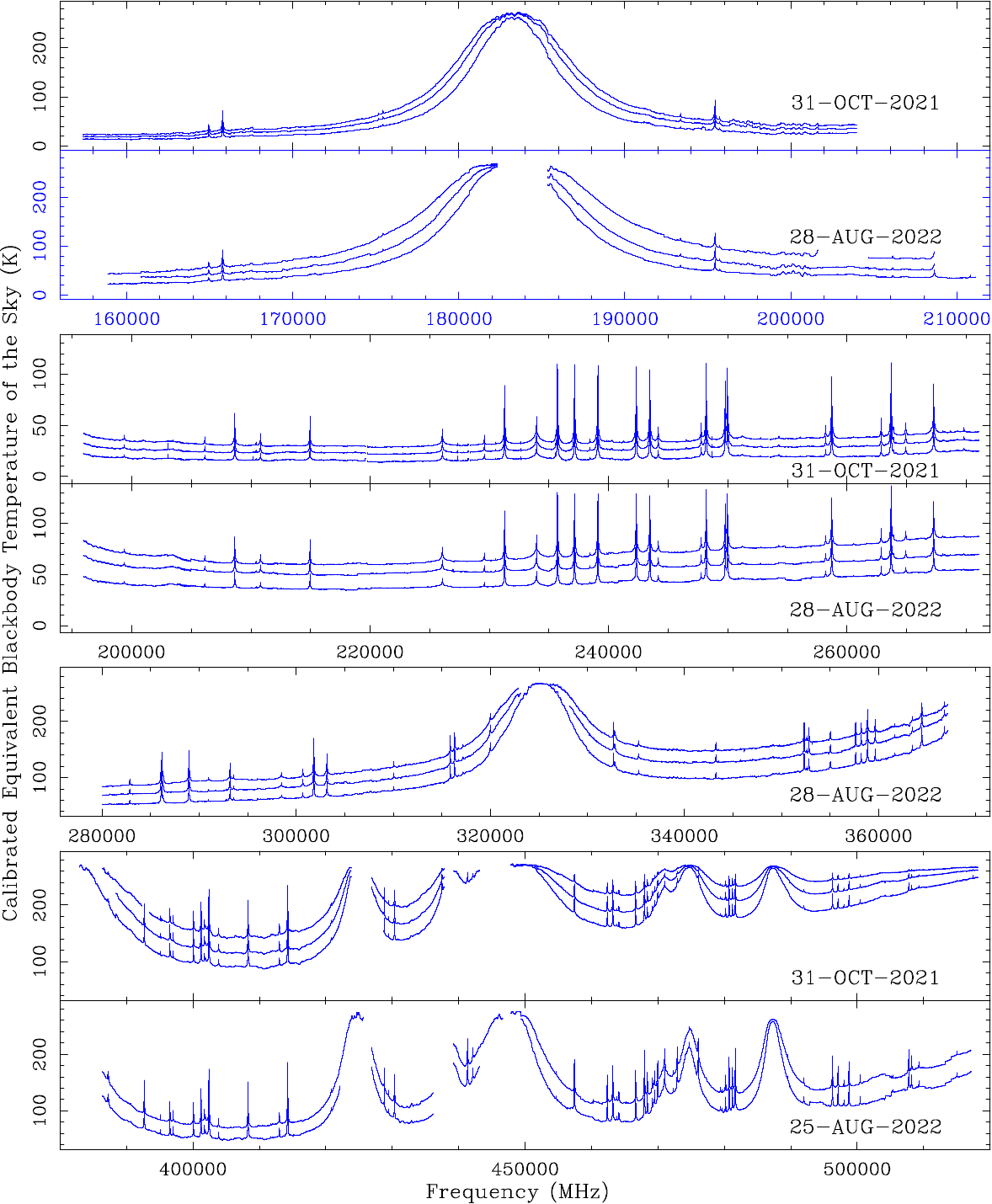}                    
	\caption{Subset of the 56 fully calibrated and cleaned atmospheric scans used
          in this study to validate CIA absorption terms in the ATM radiative transfer model. {The several
            curves in each panel correspond to observations at different air mass values for each date as
            listed in Table \ref{tb:retrievals}. The air mass increases from lower to upper equivalent
            blackbody temperature curves.}}
        \label{fg:cia_datasets}
\end{figure*}

With respect to our previous \cite{Pardo2022} work, based on one particular observing run,
several bugs in the calibration software have been fixed (replaced scalar hot/cold load
handling with Planck spectrum; fixed a mathematical bug concerning forward
efficiency division), and the sky coupling in the
calibration routine has been updated to 0.985 instead of 0.950 in that work, giving now
much more realistic outputs in the centre of opaque atmospheric lines in the observed bands,
where we know that T$_{\rm EBB}$ should reach a value equal or very close to the ambient
temperature, or the temperature of the bottom atmospheric layer. This change of sky coupling
has been independently verified using other observations like skydips, low elevation / high
opacity scans, and checking the bogus negative sky temperatures that resulted from
using 0.950. All these fixes have solved the data issues
that motivated forced corrections and alternative retrieval strategies in \cite{Pardo2022}. Although
some instrumental and/or calibration problems may remain, they are minor and
not systemic so that the overall analysis that we perform in this paper with the 56 final
spectra makes sense.

The APEX calibration software has been updated. The bug fixes change absolute values like T$_{sky}$, but do
not affect astrophysical observations taken in differential mode with the sky as off position. However, the
change in sky coupling does have a small impact on line intensities since T$_{sky}$ is slightly different and can thus lead to a different opacity.

\section{Observing runs and results}
\label{sct:obs}


In order to conduct the necessary observing campaigns, we applied for APEX telescope time via a ``calibration''
proposal in 2019, and again in 2022 for an extension of the same programme (105.2084.001). In total, we obtained around 40 hours of telescope time that were
divided into several observing sessions focusing on different atmospheric windows, seasons, time of the day, 
and overall atmospheric conditions, in particular precipitable water vapour columns (PWVC).

The observational
part of this study was affected by the global COVID-19
pandemic that prevented some panned trips to the observatory for conducting these highly non-standard
observations on-site. Nevertheless, the team managed to have a series of on-line meetings to refine the observing
strategy after each run and to discuss different technical aspects of these observations and the
progressing results. Normal operations were resumed in 2022, and J.R. Pardo was also able to travel to the telescope 
August-September 2022 for scheduled observations.

We focused our previous \cite{Pardo2022} paper on the observations conducted on December 6$^{th}$ 2020,
under very dry atmospheric conditions. It is time now to perform overall analysis using all the data obtained
in 11 different runs between October 21$^{st}$ 2020 and September 3$^{rd}$ 2022. The first calibration of the
December 6$^{th}$ 2020 data revealed calibration problems that took a long time to be solved. Many discussions
were necessary to identify different issues and to try to implement satisfactory solutions.
The main difference between using the telescope for atmospheric measurements and for astronomical observations is
that in the first case we have to achieve an accurate absolute calibration using two reference loads, and in the
second case we work in differential mode between the astronomical target and the sky around it. In addition, the strong
signal from the atmosphere can cause baseline problems and we cannot subtract any baseline.

The final spectra used in this work are fully calibrated in absolute terms after all the above mentioned solutions
were implemented (see Section \ref{sct:cal}).

Within each observing date, spectra at different airmasses were recorded in order to check for consistency
in the PWVC retrievals. However, two different strategies were followed for those skydips as several tunings
are necessary to cover all frequencies reachable by one particular receiver.

\begin{figure}[hbt!] 
 	\centering 
	\includegraphics[width=\columnwidth]{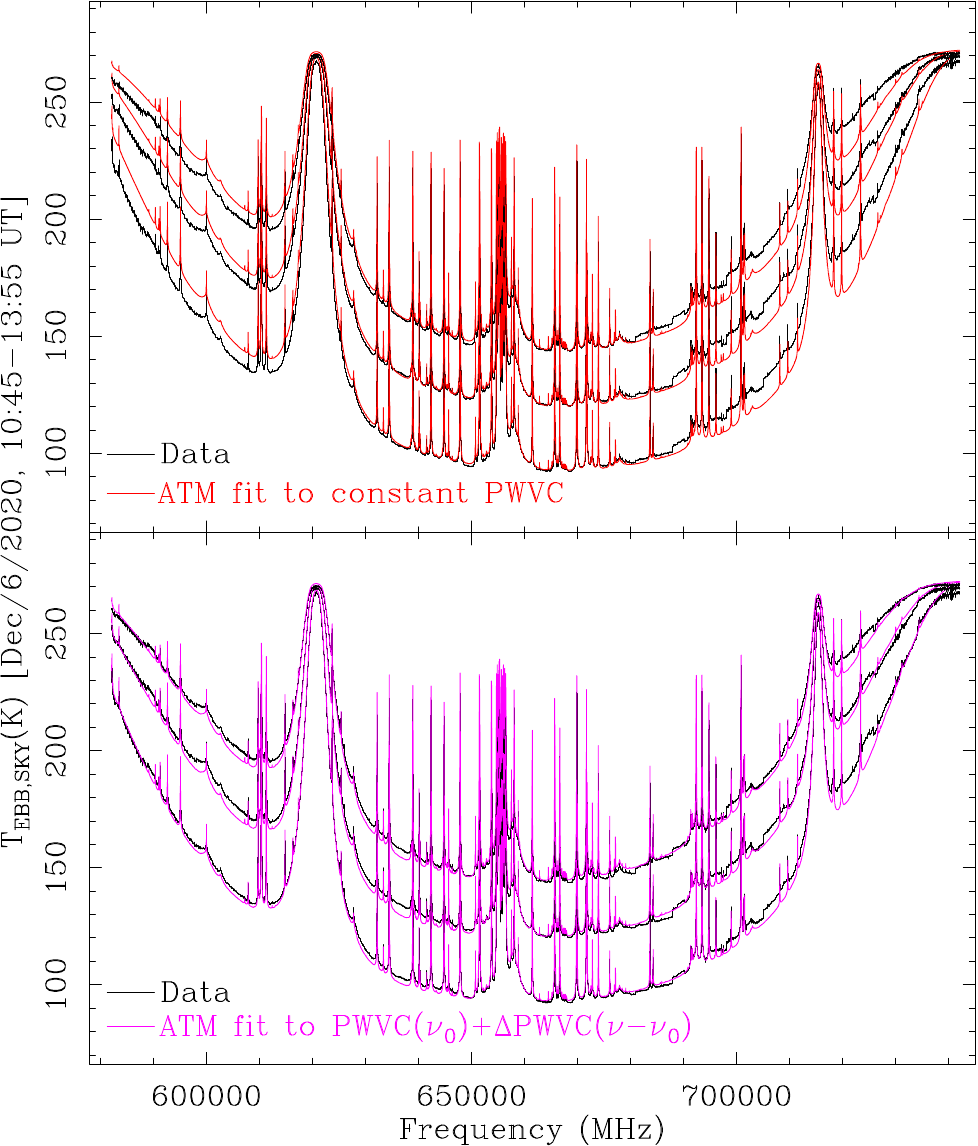}                                   
	\caption{ATM model fit to the skydip data obtained with APEX on Dec. 6$^{th}$ 2020 using
          the SEPIA660 receiver. Only three airmasses (1.0, 1.5 and 2.0) are plotted here for clarity
          (scans at 1.25 and 1.75 airmass were also taken).
          Since the observations took over three hours to complete, the PWVC changed significantly and it
        was necessary to fit an average value plus a slope for it, with a great improvement in the agreement.}
        \label{fg:20201206660}
\end{figure}

In some cases, the tuning was done and the telescope moved to the different airmasses to take data in all of
them before moving to the next frequency tuning. This implies that a quite long time was necessary to complete
the frequency coverage in one band for all the airmasses, giving enough time to the PWVC to change significantly.
This is the case of the SEPIA660 skydip of 5 airmasses (1.0, 1.25, 1.5, 1.75, and 2.0) taken on Dec. 6$^{th}$ 2020.
It took 3 hours and 10 minutes to complete it. The ATM model gives an average PWVC from fitting those data of $\sim$0.34 mm,
in good agreement with the average value retrieved from the water vapour radiometer (WVR). However, according also to
the WVR data there was an PWVC increase from the
beginning to the end of those observations of about 10\% the average value. Introducing this slope in the inputs
of ATM significantly reduces the
fit residual, giving support to the analysis (see Figure \ref{fg:20201206660}).

In later cases,
we adjusted the observing setup to keep the environmental conditions as constant as possible
for a given airmass. The elevation was kept constant, passing through all frequency tunings before moving to a different
elevation or airmass. The number of airmasses was reduced to three (1.0, 1.5 and 2.0).
In those cases, a complete frequency coverage of a particular band for a given elevation takes
a shorter time, thus reducing the chances of strong PWVC changes. However, the PWVC evolution can show up from one
airmass to another.

Not all the observing runs provided useful results due to technical problems, errors in the observing procedure
due to its highly non-standard nature, or too large fluctuations in the atmospheric conditions. A total
of 56 useful spectra providing a reasonably good coverage of frequencies
and atmospheric conditions, could be used for the goals of this study. The most relevant information about
these observing runs is summarised in Table \ref{tb:retrievals}, such as receiver used, valid frequency range, airmasses,
date, UT range, and average atmospheric P/T conditions from the weather station.

\section{Analysis}
\label{sct:analysis}
The main goal of this atmospheric study with APEX is to check the accuracy of the
current ATM model at all frequencies
covered by the observations presented in the previous section.

The model
relies upon a description of the different opacity terms (lines + CIA). Additionally,
the model uses a priori vertical profiles of pressure, temperature, and molecular abundances
based on simple assumptions from the available data provided by the weather station and
the water vapour radiometer. Specifically, we focus in this paper on the validation of
the CIA absorption terms (in units m$^{-1}$) described in the model as follows \citep{Pardo2001b} 
as a function of the frequency $\nu$ in GHz, the local temperature T in K, the water partial pressure
 $P_{H_2O}$ in mb, and the dry partial pressure ($P_{dry}$=$P$-$P_{H_2O}$ in mb):

\begin{eqnarray} \nonumber
  CIA_{(O_2-H_2O)+(N_2-H_2O)}= \\
  \;\;\;\;\;\;\;\;\;\;\;\;\;\;
  0.0315\cdot\Bigl(\frac{\nu}{225}\Bigr)^{2} 
\cdot\Bigl[\frac{P_{H_2O}}{1013}\cdot\frac{P_{dry}}{1013}\Bigr]\cdot 
\Bigl(\frac{300}{T}\Bigr)^{3}
\label{continuo-h2o}
\end{eqnarray}

\begin{eqnarray} \nonumber
  CIA_{(N_2-N_2)+(N_2-O_2)+(O_2-O_2)}=  \\
    \;\;\;\;\;\;\;\;\;\;\;\;\;\;
 2.612\cdot10^{-6}\cdot\Bigl(\frac{\nu}{225}\Bigr)^{2}\cdot
  \Bigl(\frac{P_{dry}}{1013}\Bigr)^{2}\cdot\Bigl(\frac{300}{T}\Bigr)^{3.5}
\label{continuo-seco}
\end{eqnarray}

\clearpage

It is necessary to add another dry term (3) due to the non-resonant relaxation (Debye)
spectrum of O$_2$ [\cite{Liebe1993}, section 2.2]. For the frequency range, observation geometry
and atmospheric
conditions explored in this work, this extra term produces an integrated contribution to the total
zenith opacity of 0.0029 (constant with frequency) which is slightly less than 6\% of the value
produced by the dry CIA according to equation 
\ref{continuo-seco} at 650 GHz. However, the importance of this constant term with respect to the
$\propto\nu^2$ dry CIA obviously increases as the frequency decreases. In our analysis, both
dry non resonant terms are included or removed simultaneously.

The integral of the (\ref{continuo-h2o}) and (\ref{continuo-seco})+(3) terms through the whole atmospheric
path results in the red and pink curves, respectively, of Figure \ref{fg:atmref}, {which are examined
  in the rest of this paper}, using the 56 SEPIA and nFLASH APEX atmospheric spectra summarised
in Table \ref{tb:retrievals} and the scans simultaneously performed the APEX WVR. This instrument
provides measurements of the atmospheric brightness temperature of the 183 GHz water line in 6
defined band-passes to spectrally characterise the emission symmetrically from the centre of
the water line (183.310 GHz) with a given bandwidth. 


From the closest to the H$_2$O line centre
outwards, the offsets of the WVR double sideband channels
are (in GHz): $\pm$0.6, $\pm$1.5, $\pm$2.5, $\pm$3.5, $\pm$5.0,
and $\pm$7.5. Their bandwidths, respectively, are (also in GHz): 0.2, 0.2, 0.2, 0.2, 0.4
and 0.5.

Once the spectral data are taken, an atmospheric radiative transfer model (ATM) 
is used to fit the observations and estimate the PWVC. The key point here
is that the
atmospheric opacity, and therefore the brightness temperature are both largely dominated
by the 183 GHz water vapour line so that the derived PWVC from the WVR channels would barely
reflect other aspects
of the model such as minor gases, CIA or far wings of other H$_2$O lines (see Figure
\ref{fg:atmref}. The percentage of the total opacity in the central three WVR channels
(within $\pm$2.6 GHz form the line centre) that can 
be attributed to the water line exceeds 95\% for 1 mm PWVC and the dry opacity represents less than
1.5\%. However, PWVC retrievals from the whole set of frequencies in the APEX spectra presented
in this work are quite different as minor gases and CIA opacity are in general much more important
in the large frequency ranges of our APEX atmospheric spectra. In general, water lines do not  
dominate the opacity at all frequencies. Therefore, a simple
exercise comparing PWVC derived from WVR data and from fitting the APEX spectra themselves, would
provide a very strong model validation tool, chiefly for CIA.

In order to perform the proposed validation, we smoothed the 56 spectra to a resolution of 9.7 MHz,
large enough for the broad atmospheric H$_2$O and O$_2$ lines and still providing several tens of
channels on narrower O$_3$, N$_2$O, CO, and other features. We fitted those spectra to retrieve
PWVC under some simple assumptions:

\begin{itemize}
\item Pressure and temperature at the ground are fixed to the average values provided
  by the weather station during the scan.
\item Tropospheric temperature lapse rate is fixed to -5.6 K/km.
\item Tropospheric water vapour scale height is fixed to 2.5 km.
\item Stratospheric and mesospheric P/T profiles are fixed to the \citealt{us76} values for
  a Tropical atmosphere, eventually displaced to match the values in the tropopause reached with
  the above assumptions.
\item The ozone profile is fixed by hand to minimise residuals near its lines, but with no true fit. 
\item Vertical distributions of other minor gases are left as in the \citealt{us76} Tropical
  atmosphere.
\end{itemize}

In a separate publication, finer fits for O$_3$ and other gases will be performed. For now, a first fit is
carried out with the full opacity breakdown in the ATM model: H$_2$O and dry atmospheric lines
below 1 THz, including all relevant isotopologues and vibrationally excited states, dry and wet
collision-induced absorption, and far wings of supra-THz water vapour lines. Then other fits are
carried out removing one or several of these opacity contributions.

\section{Discussion}
\label{sct:discussion}

In recent years, a number of publications have addressed the question of the non-resonant foreign wet 
and dry CIA in the atmosphere from theoretical calculations or laboratory experiments under well
controlled conditions (\citealt{Boi2003}, \citealt{Podo2008}, \citealt{Tret2015}, \citealt{Ser2020}).
Our observations, of course, are under less controlled conditions and we detect the radiation after
its propagation through the whole atmosphere to our detectors. Therefore, we cannot address the
fine details given in the above-cited publications. However, the description of these opacity
terms in ATM for the conditions of millimetre and submillimetre ground-based observatories can
be validated with our data, even with the experimental data {scatter} that we can expect in this
3 year study. Even so, this validation is a big step forward in the state of the art of
atmospheric models used by the observatories.

\begin{figure*}[ht!] 
 	\centering 
	\includegraphics[width=1.00\textwidth]{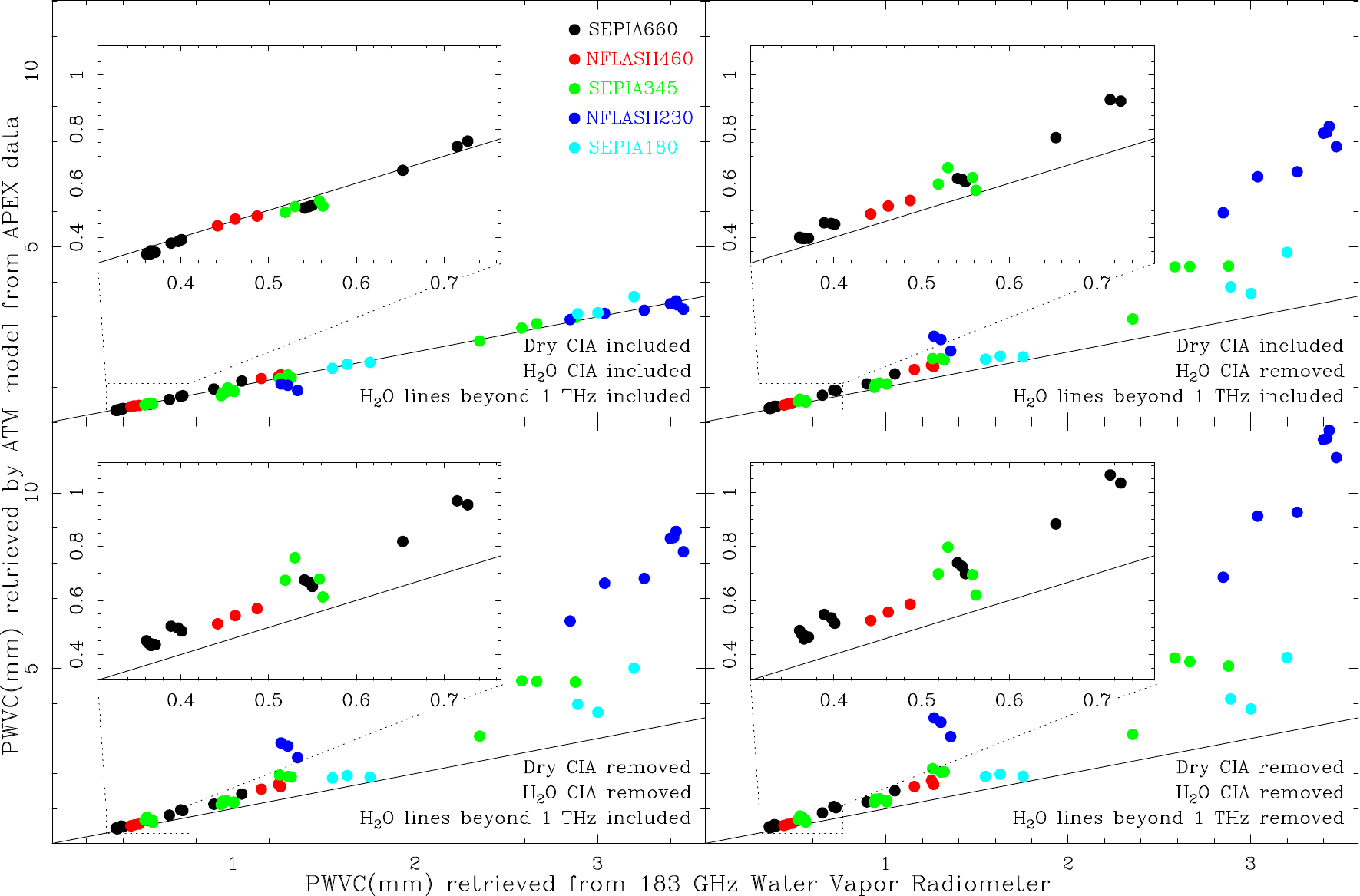}     
	\caption{PWVC provided by the water vapour radiometer (average over the scan) versus the one
          derived from ATM fits of APEX spectra under various scenarios. The inset in each panel
          corresponds to a zoom in the area marked by the dotted box. {The black solid line in
          each panel, and in the insets, marks the bisector/diagonal or equal PWVC values derived from the two methods.}}
        \label{fg:corr}
\end{figure*}

Following the procedure explained in the previous section we have plotted,
for each one of the 56 APEX spectra, PWVC fit results against
the temporal average of the WVR-based PWVC. Although experimental scatter should be expected, if
the model has a correct CIA description, an alignment around the diagonal line should be found. On the
contrary, a misalignment should appear if the full model is inconsistent, or if different terms of
the correct model are removed. Figure \ref{fg:corr} shows the results of this exercise. The top-left
panel, corresponding to the full ATM model shows a relatively good agreement for the whole PWVC range of
$\sim$ 0.35 mm to 3.5 mm, or one order of magnitude. An inset provides a closer look into the 0.3-0.8 mm PWVC
range. There are three dark blue dots around 1.0-1.3 mm corresponding to a three airmass skydip taken
with nFLASH230 on Oct. 31$^{st}$ 2021 that are a bit off the line. Since the nFLASH230 band does not include
any strong water line, it is the most sensitive to possible calibration issues. Nevertheless, we have decided
to keep those spectra in the analysis as they provide significant information on all other panels.

The disagreement
between the dots in the top-left panel and the perfect diagonal line is only 6.25\%, a value that {seems} 
compatible with the expected experimental {scatter}. {There are three blue data points (NFLASH230 receiver)
around 1.2-1.4 mm PWVC that show the largest deviation from the diagonal line. Due to the very low atmospheric opacity
in the frequency range of the NFLASH230 receiver, deviations like this can somehow be expected as the PWVC retrieval
from that frequency range is more uncertain than at higher frequencies. However, due to the large number of other data
points, the conclusions do
not change much. In fact, other blue data points in this panel at PWVC > 2.8 mm, for which the PWVC retrieval
from the NFLASH230 spectra is less uncertain, show much less devitation from the diagonal.}

The top-right panel of Figure \ref{fg:corr} shows that, as expected, the ``foreign'' wet collision-induced
absorption due to both O$_2$-H$_2$O and N$_2$-H$_2$O
collision mechanisms, is a relevant element in the model as its removal produces a large
disagreement in that panel (+55.79\% average difference of the vertical axis values with respect to
the horizontal axis ones, retrieved from the WVR, taken as a reference as explained in
Section \ref{sct:analysis}). The largest disagreement
is in the dark blue dots that correspond to nFLASH230 spectra in a frequency region where, as said before,
``foreign'' wet CIA is the dominant opacity term. On top of that, there is a general trend of more disagreement
for wetter situations, and this can be also seen in SEPIA180 or SEPIA345 because these spectra cover
not only the strong 183 GHz or 325 GHz water lines, but
a significant range of ``window'' frequencies where, in fact, ``foreign'' wet CIA opacity has a large 
share to the total opacity. 

In general, the dry CIA (N$_2$-N$_2$, N$_2$-O$_2$, O$_2$-O$_2$ collisions)
is much more constant and weaker (except for very dry situations) than the wet CIA. Therefore its effect
is much more limited, as shown by the plot in the bottom-left panel where we have removed the dry CIA term
from the model in addition to the ``foreign'' wet CIA. The overall difference from the diagonal line increases
now to 63.10\%.

Finally, removing the far wings of H$_2$O lines centred above 1 THz also adds to the disagreement (bottom-right
panel in Figure \ref{fg:corr}) that increases to 89.60\%.

These results are quite conclusive: The ATM model needs all the original opacity terms to provide consistent
results at least for the atmospheric conditions corresponding to high and dry millimetre and submillimetre
observatories, as it was the case for Mauna Kea and now Chajnantor. Based on these results we would not
suggest any changes in the model that may affect broadband opacity terms.

\section{Summary and conclusions}
\label{sct:concl}
We have studied the CIA terms in the atmospheric model ATM for the typical atmospheric conditions
reigning at the Chajnantor Plateau with one order of magnitude range in the PWVC, using
well absolutely calibrated spectra covering full frequency ranges for five receivers
from $\sim$ 157 to 742
GHz. The analysis reveals that all CIA terms derived $\sim$20-25 years earlier from atmospheric scans
with the Caltech Submillimetre Observatory in Mauna Kea seem to be correct to the extent of the
experimental {scatter} with APEX at Chajnantor, and should remain unchanged in the model. Further
measurements are advised with the nFLASH230 receiver due to its special sensitivity to the CIA terms.
Completing the study with more data beyond 752 GHz, into the next submillimetre window, and
below $\sim$ 170 GHz (including in higher PWVC conditions that can be used for astronomical
observations at such frequencies), is also
desirable. The results of this study strongly depend on a correct absolute calibration and, for
that reason, a long time has been devoted to understand and refine it as much as possible
resulting on a benefit for the APEX telescope.

\begin{acknowledgements}
The authors thank ESO for funding this work through contract CFP/ESO/19/25417/HNHE. JRP also thanks
ESO for funding the implementation of ATM in the ALMA software under contract 14977/ESO/07/15694/YWE. 
Additional support has also been provided by EUMETSAT under contract EUM/20/4600002477. {Part of this
work was carried out at the Jet Propulsion Laboratory, California Institute of Technology, under
a contract with NASA (80NM0018D0004).}

\end{acknowledgements}

\end{document}